\documentclass[preprintnumbers,a4paper,twocolumn,floats,aps]{revtex4}
\usepackage{graphicx}
\usepackage{amsmath,amssymb,amsfonts}
\usepackage[dvips]{hyperref}
\begin{document}
\title{On the DAMA and CoGeNT Modulations}

\preprint{OUTP-11-40-P}

\author{Mads T. Frandsen}
\email{m.frandsen1@physics.ox.ac.uk}
\author{Felix Kahlhoefer}
\author{John March-Russell}
\author{Christopher McCabe}
\author{Matthew McCullough}
\author{Kai Schmidt-Hoberg}

\affiliation{Rudolf Peierls Centre for Theoretical Physics, University of Oxford, Oxford OX1 3NP, United Kingdom}

\begin{abstract} 
DAMA observes an annual modulation in their event rate, as might be expected from dark matter scatterings, while CoGeNT has reported evidence for a similar modulation. The simplest interpretation of these findings in terms of dark matter--nucleus scatterings is excluded by other direct detection experiments. We consider the robustness of these exclusions with respect to assumptions regarding the scattering and find that isospin-violating inelastic dark matter helps alleviate this tension and allows marginal compatibility between experiments. Isospin-violation can significantly weaken the XENON constraints, while inelasticity enhances the annual modulation fraction of the signal, bringing the CoGeNT and CDMS results into better agreement.
 \end{abstract}

\date{\today}
\maketitle

\section{Introduction} 

Gravitational effects on astrophysical scales give convincing evidence for a large abundance of particle dark matter (DM) in the universe, an observation which is strongly supported by measurements of the cosmic microwave background anisotropy \cite{Jarosik:2010iu}. In contrast, very little is known about the properties of the DM particle. One of the most promising strategies for its identification are direct detection experiments, aiming to observe the scattering of DM particles off target nuclei. While the recoil energy spectrum has no features that allow for an unambiguous identification of a DM signal, a characteristic annual modulation of the differential event rate is expected due to the motion of the Earth relative to the Galactic halo \cite{Drukier:1986tm, Freese:1987wu}.

Two DM direct detection experiments, namely DAMA \cite{Bernabei:2010mq} and CoGeNT \cite{Aalseth:2010vx, Aalseth:2011wp}, have published data or presented preliminary results demonstrating evidence for such an annual modulation. The combined results from DAMA/NaI and DAMA/LIBRA have a statistical significance exceeding $8\sigma$. CoGeNT has taken data for over a year and also observes an annual modulation with a significance of $2.8\sigma$ \cite{Aalseth:2011wp}. The simplest explanation of these experiments in terms of DM is spin-independent elastic scattering on both protons and neutrons of a light, $\mathcal{O}(10)$ GeV DM particle \cite{Bottino:2003cz, Kelso:2010sj}. However, this explanation is strongly disfavoured by other null results, most notably by the CDMS \cite{Ahmed:2010wy}, XENON10 \cite{Angle:2011th} and XENON100 \cite{Aprile:2011hi} experiments.

Various proposals towards reconciling all experiments have been put forward. First of all, doubts have been raised concerning experimental details, such as the proper calibration of the nuclear recoil energy scale and the correct assessment of the various quenching factors (see e.g.~\cite{Collar:2010ht}). Secondly, significant astrophysical uncertainties are present in all analyses, which may considerably change the implications of the data \cite{MarchRussell:2008dy,McCabe:2010zh}. Finally, it may well be that the DM-atom interaction is not solely with the nucleus or that the DM-nucleus interaction is more complicated than generally assumed.

In this paper we will primarily focus on the third option in the context of nuclear recoils. Of particular interest is the combination of two effects that have been recently considered as possibilities to alleviate the tension between the various experiments: Isospin-violating dark matter (IVDM) \cite{Giuliani:2005my, Chang:2010yk, Feng:2011vu} and inelastic dark matter (iDM) \cite{TuckerSmith:2001hy,TuckerSmith:2004jv,Chang:2008gd, SchmidtHoberg:2009gn,Cline:2010kv}. In fact, if one assumes that DM scatters differently on protons and neutrons, then for a particular choice of the proton to neutron scattering fractions, it is possible to weaken the limits from XENON, which would otherwise give the strongest constraints \footnote{Note that it is not possible to completely suppress scattering on xenon due to the presence of different isotopes, as emphasised in \cite{ Feng:2011vu}.}. On the other hand, inelastic scattering of the DM particle can enhance the annual modulation signal and reduce the tension between CDMS and CoGeNT, which cannot be reduced by isospin-violating couplings.

We only take the annual modulation of the signal measured by DAMA and CoGeNT as a hint of DM. We make the important assumption that the background does not modulate, and that the modulation is entirely due to DM interactions. We do not fit to the unmodulated spectrum as it is very difficult to interpret an exponentially falling energy spectrum in the presence of an unknown background in terms of a DM particle. Rather we can use the unmodulated recoil spectra for the sole purpose of calculating exclusion limits to constrain the parameter region consistent with annual modulation.

\section{Direct Detection of Dark Matter}
\label{sec:overview}

The differential event rate for DM-nucleus scattering is (see e.g.~\cite{Lewin:1995rx})
\begin{equation}\label{dRdE1}
\frac{dR}{dE_R}=N_T\frac{\rho_{\chi}}{m_{\chi}}\int^\infty_{v_{\text{min}}} v f_{\text{local}}(\vec{v},t) \frac{d \sigma}{d E_R}d^3v ~~,
\end{equation}
with $N_T$ the number of target nuclei per unit mass, $m_\chi$ the DM mass and $\rho_\chi$ the local DM density. $f_{\text{local}}$ is the local DM velocity distribution, which we assume is an isotropic Gaussian distribution with velocity dispersion $v_0=220\text{ km/s}$, truncated at the galactic escape velocity, which we take as $v_{\text{esc}}=544\text{ km/s}$.

In a direct detection experiment the minimum speed that an incident DM particle must have in order to transfer an energy $E_R$ to a recoiling nucleus is \cite{TuckerSmith:2001hy}
\begin{equation}
\label{eqn:idmvelo}
v_\text{min} = \frac{1}{\sqrt{2 m_N E_R}}\left( \frac{m_N E_R}{\mu} + \delta \right) \;.
\end{equation}  
$m_N$ is the mass of the target nucleus, $\mu$ is the reduced mass of the DM-nucleus system and $\delta$ is the mass difference between the incoming and outgoing DM particle. 

For coherent spin-independent DM nucleus scattering, the DM-nucleus differential cross-section is
\begin{equation}\label{dsigmadE}
\frac{d \sigma}{d E_R}=\frac{1}{2 v^2}\frac{m_N \sigma_n}{\mu^2_{\chi n}}{\frac{(f_p Z + f_n (A-Z))^2}{{f_n}^2}} F^2(E_R) ~~,
\end{equation}
where $\mu_{\chi n }$ is the DM-nucleon reduced mass and $\sigma_n$ is the DM-neutron cross-section at zero momentum transfer in the elastic limit. $f_{n,p}$ are the effective coherent couplings to the neutron and proton respectively, while  $A$ and $Z$ denote the nucleon and proton numbers of the target nucleus. The nuclear form factor $F(E_R)$ encodes the loss of coherence as the momentum transfer deviates from zero. 

For DAMA, we fit to the latest data \cite{Bernabei:2010mq} and take the sodium quenching factor to be $Q_{\text{Na}}=0.3$. In one case we consider $Q_{\text{Na}}=0.43$, within the range $Q_{\text{Na}} = 0.3 \pm 0.13$, deemed representative of experimental uncertainties in \cite{Hooper:2010uy}. We do not include channeling in our calculations \cite{Bozorgnia:2010xy}. We find that scattering off of sodium dominates that of iodine in the low mass DM region that we consider. We use the detector resolution from \cite{Bernabei:1998rv} for DAMA/NaI and \cite{Bernabei:2008yh} for DAMA/LIBRA and weight them appropriately.

\begin{figure*}[!th]
{
\includegraphics[width=1.0\columnwidth]{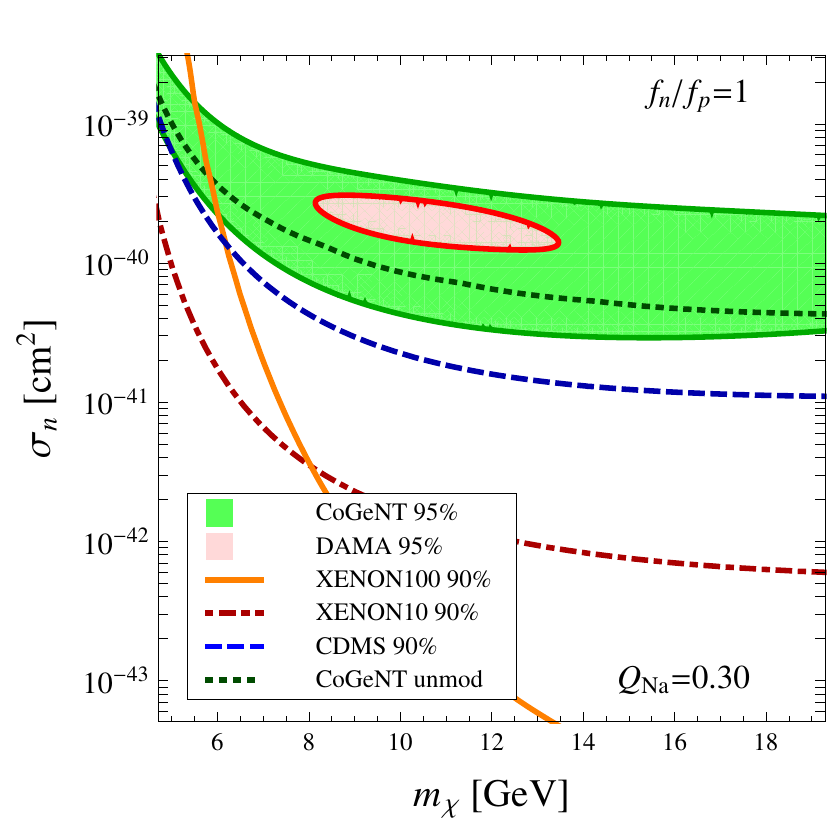}
\includegraphics[width=1.0\columnwidth]{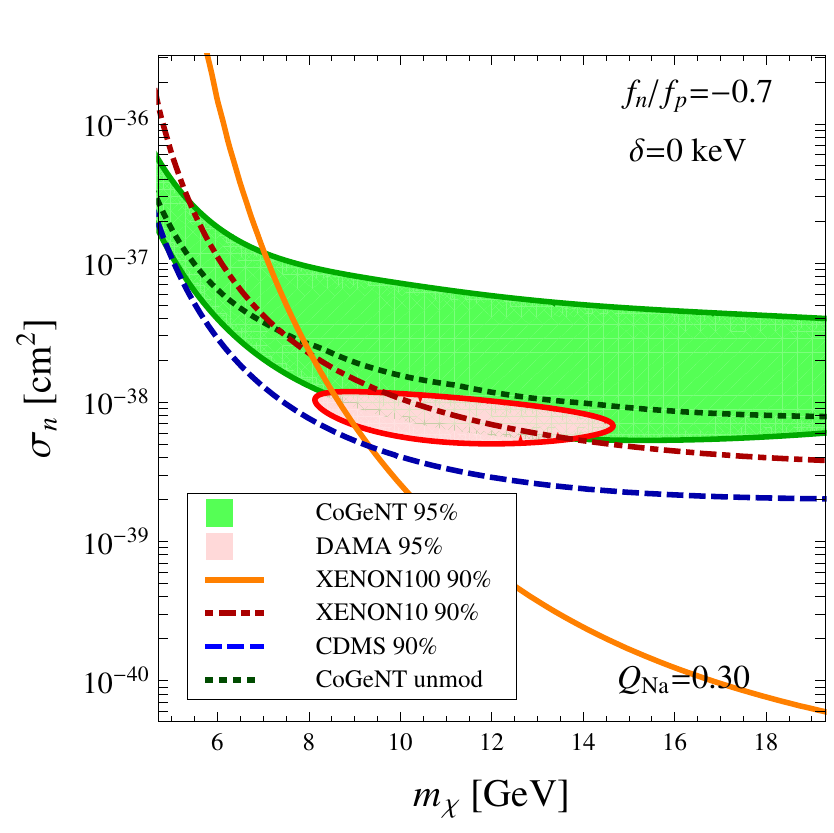}
\includegraphics[width=1.0\columnwidth]{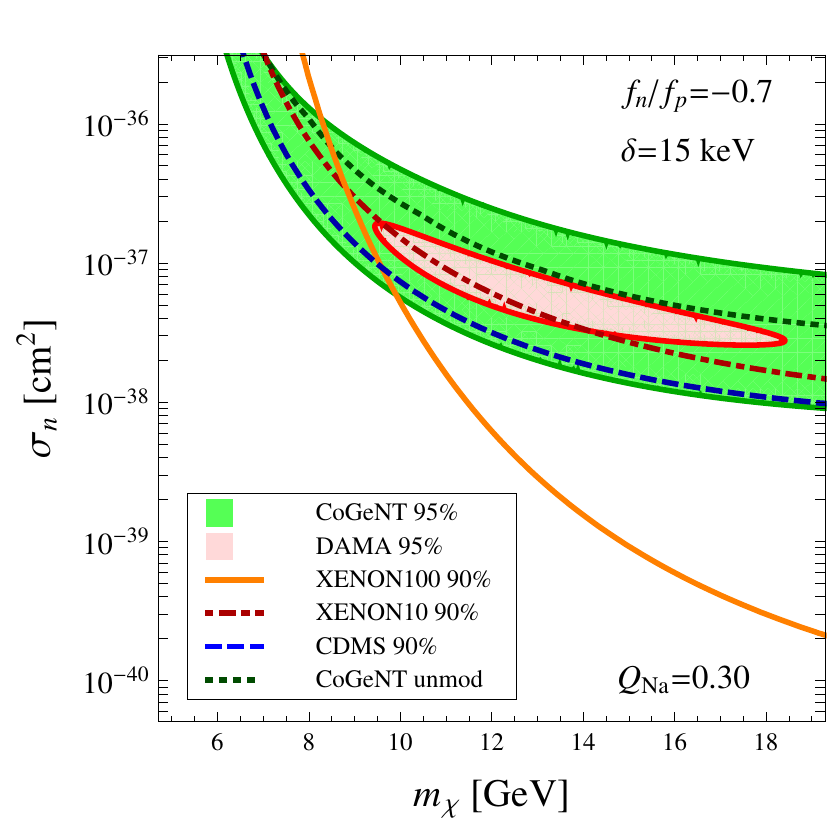}
\includegraphics[width=1.0\columnwidth]{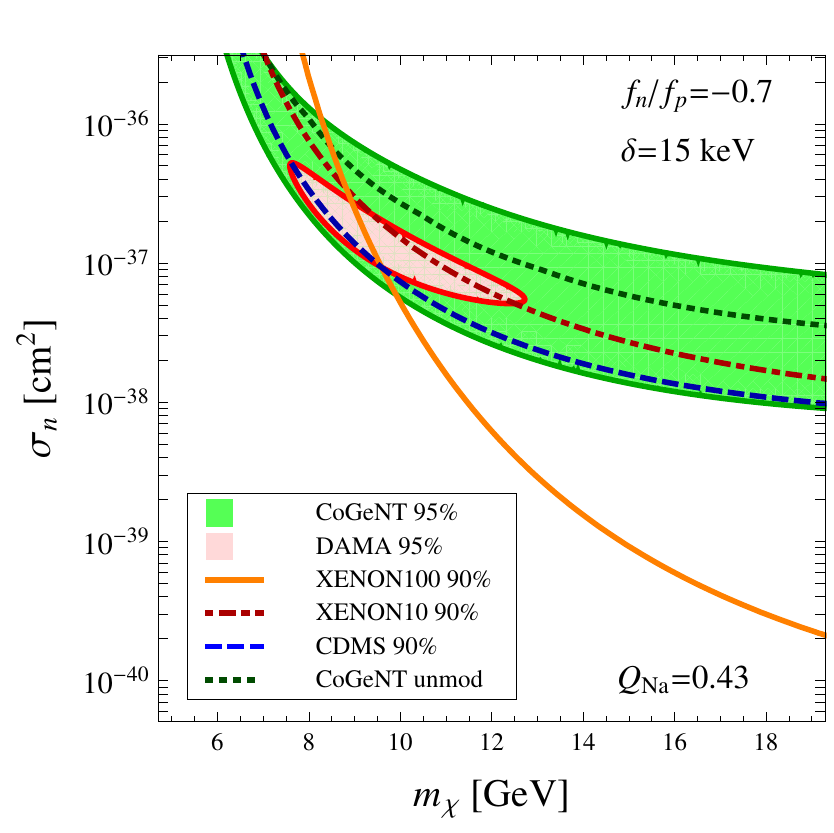}
}
\caption{Best-fit parameter regions for DAMA and CoGeNT (coloured regions) as well as exclusion limits from XENON10, XENON100, CDMS II and the unmodulated CoGeNT signal. In the upper panels, $\delta = 0$ keV and $f_n / f_p = 1$ (left) and $f_n / f_p = -0.7$ (right).  The lower panels have $\delta = 15$ keV and $f_n / f_p = -0.7$ (left and right).  In the lower right panel we have taken the sodium quenching factor to be $Q_{\text{Na}} = 0.43$. In the other panels we take the standard value $Q_{\text{Na}} = 0.3$. Notice the upper left panel uses a different scale for $\sigma_n$. It is clear that employing the IVDM and iDM mechanisms significantly weakens the constraints from null searches, and allows for a small region of agreement when the sodium quenching factor is varied within a reasonable range.}
\label{fig1}
\end{figure*}

For the CoGeNT parameter regions shown in Fig.~\ref{fig1}, we use modulation amplitudes of $1.20 \pm 0.65$, $0.54 \pm 0.19$ and $0.08 \pm 0.16$ events/kg/day/keVee for $0.5-0.9$, $0.9-3.0$ and $3.0 - 4.5$ keVee respectively. We use the detector resolution and efficiency given in \cite{Aalseth:2010vx, Aalseth:2011wp} and use a quenching factor for germanium of $0.2$.

A DM explanation of a signal in DAMA and CoGeNT must be confronted with the exclusion limits obtained by the XENON and CDMS collaborations. While heavy DM is most strongly constrained by the recent results of XENON100 \cite{Aprile:2011hi}, an even stronger bound for low mass DM can be obtained from a dedicated low threshold analysis of the XENON10 data \cite{Angle:2011th}. For the XENON experiments we assume a detector resolution dominated by Poisson fluctuations and take $\mathcal{L}_{\text{eff}}$ and $\mathcal{Q}_y$ from the respective papers. For the XENON10 data, we include the S2 width cut and assume the absence of fluctuations at the low energy threshold (i.e.~impose a cut-off).  

The ratio $f_n/f_p$ depends on the underlying model of DM and generally differs from one.  Eqn.\ \ref{dsigmadE} implies that if the DM scattering satisfies $f_n/f_p=Z/(Z-A)$ for a given nuclear isotope, this isotope would then give no constraint.  In practice, experiments consist of targets with more than one isotope, so they can still yield some constraints. As considered in \cite{Chang:2010yk,Feng:2011vu} the choice  $f_n/f_p\sim -0.7$ reduces the sensitivity of xenon experiments by three orders of magnitude.

The most restrictive limit from CDMS results from a low-energy analysis of the CDMS II data from the Soudan Underground Laboratory \cite{Ahmed:2010wy}. In this paper we only include the detector with the best ionisation resolution (T1Z5), treating all observed events as potential DM signals. We find that the constraint obtained with this simplification agrees well with the limit published by CDMS. Since both CoGeNT and CDMS are germanium target experiments they cannot be reconciled using isospin-violating couplings. Inelastic scattering increases the modulation fraction, and hence reduces the tension between CDMS and CoGeNT.  However, for scattering on sodium in DAMA the tension between DAMA and CDMS is not reduced as germanium is greater in mass. 

In Fig.~\ref{fig1} we show how the combination of the iDM and IVDM mechanisms can reduce but not completely resolve the apparent conflict between the annual modulation observed in both CoGeNT and DAMA with exclusion limits from other detectors. Any alteration of these limits due to uncertainties in low-energy efficiencies, astrophysical parameters, quenching factors, or inclusion of channelling could weaken the limits sufficiently to allow a DM interpretation of the CoGeNT and DAMA annual modulation. As a demonstration, we consider a larger value of the sodium quenching factor and find better agreement as shown in the lower right panel of Fig.~\ref{fig1}. In fact, it should be noted that our 95\% best-fit and 90\% exclusion regions do not take into account uncertainties in all such astrophysical and experimental parameters, and therefore are not truly 90\% regions.

\section{Future constraints from direct detection experiments}
\label{sec:future}

Even with $f_n/f_p=-0.7$, the XENON100 limit remains strong and cannot be suppressed much further due to the different isotopes present in natural liquid xenon. Consequently, additional data from XENON100, as well as a dedicated S2 only analysis with lower threshold may provide strong constraints for isospin-violating inelastic DM. The CDMS collaboration has also collected data from detectors with a silicon target. From the preliminary data presented in \cite{Filippini:2008zz}, we find a limit which is more constraining than the CDMS germanium low threshold analysis. We have not included this in Fig.~\ref{fig1} because of uncertainties related to the calibration of the silicon nuclear recoil energy scale \cite{Filippini:2008zz, Hooper:2010uy, JPFilippini2011}. We encourage the CDMS collaboration to perform a dedicated analysis of the Soudan silicon data.

Unpublished results from the CRESST experiment \cite{cresst} suggest an excess in the oxygen band of $32$ events over an estimated background of $8.7$ events.  Light DM-nuclei scattering can offer an explanation of this excess, and the DM mass and DM-nucleon cross-section required take values close to those required for an explanation of the DAMA and CoGeNT observations.  An approved analysis of the data taken by CRESST could have a significant impact on the preferred region in our scenario.

The CRESST experiment also offers the potential to test the inelastic nature of the DM because of the presence of both tungsten and oxygen in the detector.  For an elastic DM explanation of the oxygen band events a signal in the tungsten band is implied.  This signal lies at lower energies, $E_R \lesssim 4$ keV, and is usually small.  However, for inelastic scattering the picture changes, as scattering rates for tungsten are enhanced in comparison to oxygen.  This enhancement is strongly dependent on the splitting, and can, for a splitting of $\delta \sim 15$ keV, enhance the ratio of tungsten-to-oxygen scattering rates at low energies by over an order of magnitude, in comparison to elastic scattering.  In addition, at $E_R \sim 4$ keV, scattering on tungsten becomes dominant over oxygen, which would lead to a sharp rise in the number of DM-like events in the  tungsten band below $E_R \lesssim 4$ keV.  This does not, however, occur for elastic scattering of light DM and thus a detector containing both light and heavy elements, such as CRESST, could potentially distinguish between the elastic and inelastic light DM scenarios in the future.

\section{Other constraints and implications for the dark sector}
\label{sec:otherconstraints}
Mono-jet~\cite{Goodman:2010yf,Goodman:2010ku,Bai:2010hh} and di-jet searches at the Tevatron constrain the DM couplings to quarks. For light DM the Tevatron limits on mono-jet production are relevant, but sensitive to the mass of the particle mediating the DM interaction with quarks \cite{Bai:2010hh}. Assuming the mediator is heavy compared to the typical momentum transfer at the Tevatron, the limits can be as strong as $\sigma_n \sim 10^{-39} {\rm \, cm^2}$. However, if the mediator is light, or if the coupling is not by a vector current, the constraints weaken. For example, a mediator of mass $M\sim10$ GeV interacting via a vector current allows $\sigma_n \sim 3\times10^{-34} {\rm \, cm^2}$ easily allowing our best fit CoGeNT and DAMA regions.

The couplings of the mediator to quarks $g_q$ are separately constrained. In the case of a vector mediator with a mass in the range $M\gtrsim 10$ GeV the constraints are $g_q \lesssim 0.1$ from meson decay measurements \cite{Aranda:1998fr} which easily satisfies the constraints from the di-jet measurements at Tevatron. The DM-neutron scattering cross-section may be written as $\sigma_n \sim f_n^2 \, g_\chi^2 \, \mu_{\chi n}^2 /(\pi M^4)$, where $g_\chi$ is the DM mediator coupling and the effective coupling constants via the vector current $\bar{q}\gamma^\mu q$, for the neutron (proton) are $f_n = g_u + 2 g_d$ ($f_p = 2 g_u + g_d$). We can then reach the required DAMA and CoGeNT cross-section $\sigma_n \sim 10^{-37} {\rm \, cm}^2$ with $g_q\sim g_\chi\sim 5\times 10^{-2}$, within the above bounds. If the light dark matter results from strong dynamics in another sector saturating the perturbativity bound $g_\chi \sim \sqrt{4\pi} $ \cite{Frandsen:2011kt} couplings to quarks as small as $g_q \sim 10^{-3}$ are allowed.

If the DM has no particle--anti-particle asymmetry, additional constraints from searches for the annihilation products apply. For example, there are constraints coming from neutrino telescopes searching for the annihilation products of DM accreted in stars. In particular, annihilations of light DM captured in the Sun are constrained by SuperKamiokande which excludes the DAMA/CoGeNT regions shown in Fig.~\ref{fig1} for DM annihilating into the $c\bar{c}, b\bar{b}, \tau\bar{\tau}, \nu\bar{\nu}, 4\tau$ channels (even with velocity suppressed annihilations) \cite{Kappl:2011kz}.  Note that also inelastic, annihilating DM is constrained by capture and annihilation in the Sun \cite{Nussinov:2009ft,Menon:2009qj} and compact stars \cite{McCullough:2010ai,Hooper:2010es} and for the small splittings we are considering the above mentioned limits still apply.  

An attractive framework which avoids the annihilation constraints is asymmetric dark matter (ADM), where a particle-antiparticle asymmetry $\eta_{\chi}=(n_\mathrm{\chi} - n_\mathrm{\bar{\chi}})/s$ in DM, similar to that in baryons $\eta_B$, provides a natural link between their observed energy densities. If some process shares or co-generates the two asymmetries ensuring $\eta_{\chi} \sim \eta_B$, then the observed cosmological DM energy density is realised for $m_{\chi} \sim 1-10$ GeV.
Consequently, ADM offers a motivation for the low DM mass favoured by the DAMA and CoGeNT data, and naturally avoids the constraints from annihilation in the Sun \cite{Kappl:2011kz}. We leave a more thorough analysis of models within this framework which realise the required isospin-violation and inelastic splitting for future work.

\section*{Acknowledgements}

MTF acknowledges a VKR Foundation Fellowship. FK is supported by the DAAD. CM and MM are supported by STFC studentships.
We thank Peter Graham, Stephen West and Martin Winkler for discussions as well as J.~P.~Filippini for clarifying the status of CDMS II-Si. JMR and KSH acknowledge support from ERC Advanced Grant BSMOXFORD 228169. We thank the European Research and Training Network ``Unification in the LHC era'' (PITN-GA-2009-237920) for partial support. 

\bibliography{Directdetection}
\bibliographystyle{ArXiv}

\end{document}